\documentclass{article}

\usepackage{amsmath}
\usepackage{arxiv}

\usepackage[utf8]{inputenc} 
\usepackage[T1]{fontenc}    
\usepackage{hyperref}       
\usepackage{url}            
\usepackage{booktabs}       
\usepackage{amsfonts}       
\usepackage{nicefrac}       
\usepackage{microtype}      
\usepackage{lipsum}		
\usepackage{graphicx}
\usepackage{natbib}
\usepackage{doi}

\usepackage{multirow}

\title{Quantifying hierarchy in scientific teams}

\author{ Fengli Xu \\
	Knowledge Lab, Department of Sociology\\
	University of Chicago\\
	Chicago, IL\\
	\texttt{fenglixu@uchicago.edu} \\
	\And
	Lingfei Wu \\
	School of Computing and Information\\
	University of Pittsburgh\\
	Pittsburgh, PA \\
	\texttt{liw105@pitt.edu} \\
	\AND
	James A. Evans \\
	Knowledge Lab, Department of Sociology\\
	University of Chicago\\
	Chicago, IL\\
	\texttt{jevans@uchicago.edu} \\
}

\renewcommand{\shorttitle}{Conceptual Hierarchy in Scientific Teams}

\hypersetup{
pdftitle={Flat Teams Drive Scientific Innovation},
pdfsubject={Science of Science},
pdfauthor={Fengli Xu, Lingfei Wu, James A. Evans},
pdfkeywords={teams, innovation, science of science, productivity, novelty, disruption},
}

\begin{document}
\maketitle

\begin{abstract}
This paper provides a detailed description of the data collection and machine learning model used in our recent PNAS paper "Flat Teams Drive Scientific Innovation"~\cite{xu2022flat}. Here, we discuss how the features of scientific publication can be used to estimate the implicit hierarchy in the corresponding author teams. Besides, we also describe the method of evaluating the impact of team hierarchy on scientific outputs. More details will be updated in this article continuously. Raw data and Readme document can be accessed in this GitHub repository~\cite{fengli_xu_2022_6569339}.
\end{abstract}


\section{Estimating the Impact of Team Hierarchy (L-ratio) on Scientific Output}

To evaluate the impact of team hierarchy (L-ratio) on the innovation performance of individual scientists, we select scientists publishing two or more papers and perform author and field fixed-effect regressions to normalize the performance differences between authors and fields of study as defined in Microsoft Academic Graph \cite{mag_data}. We include L-ratio as an independent variable to predict six dependent variables, including novelty, developmental index, the productivity of lead authors, the productivity of support authors, short-term citation impact, and long-term citation impact. To control for possible confounders, we also include team size \cite{Wu2019-al}, the mean, standard deviation and max value of career age \cite{blau2017us}, whether the work is supported by funding agencies, and the number and award amount of received funding as independent variables in predicting each dependent variable. We find L-ratio continues to be statistically significant across all regressions. L-ratio explains the most additional variance in predicting the increased developmental index (167\%), increased short-term citations (121\%), and decreased novelty (17\%). The estimated regression coefficients also allow us to compare the effect of different variables. This suggests, for example, that changing team structure (engaging a support author as a lead author) may be more effective than changing team size in maximizing novelty.

\section{Quantifying the Novelty of Research Papers}

The novelty metric (Fig.2a) quantifies to what extent a paper links topic keywords that rarely appear together. Our metric is designed to extend the Uzzi score of reference novelty. Brian Uzzi and colleagues created a prominent score that captures how a paper deviates from the norm of science by building on “atypical” references, where a pair of journals are determined to be “atypical (z < 0)” if they are less likely than random to be co-cited in the existing literature \cite{Uzzi2013-wd}. 

Here we leverage the established scientific taxonomy \cite{shen2018web} and measure keyword novelty rather than reference atypicality, as the former more directly reflects the expanded cognitive extent of a focal paper \cite{stasa2015quantifying}. The original Markov Chain Monte Carlo (MCMC) algorithm used to calculate atypicality, however, cannot practically calculate the pairwise atypicality between 550k topic keywords. We overcome this technical challenge by reformulating atypicality as distance between keywords in a knowledge space inferred from embedding techniques \cite{lin2022new, mikolov2013distributed}. This reformulation is achieved by noting that atypicality is deeply related to a common measure in information science, the Pointwise mutual information (PMI) between two items, and that PMI is formally equivalent to the inner product of two vectors representing items within a latent semantic space \cite{levy2014neural}.

Specifically, we apply the skip-gram word2vec model \cite{mikolov2013distributed} to learn the vector representation of topic keywords from their co-occurrence in research papers. The novelty $N$ of a paper is computed as,

\begin{align*}
N = \mathbf{Mean}_{i,j \in K}(- Emb_{in-i} \cdot Emb_{out-j})
\end{align*}
where $K$ is the keyword list of the given paper, $Emb_{in-i}$ is the item embedding of keyword $i$, and $Emb_{out-j}$ is the context embedding of keyword $j$. The negative inner product of the item and context embedding approximate atypicality as discussed above. We conducted multiple experiments using datasets across different years -- 1970, 1985, and 2000 -- to verify the approximation of the inner product to atypicality. The Pearson correlation coefficients between these two variables consistently lay between 0.74 and 0.75 with a $p$-value smaller than 0.001.

Higher $N$ means the paper has more surprising combinations of topic keywords. For example, \cite{hidalgo2007product} pioneered the use of complex network models in analyzing country product exporting behaviors and \cite{jeong2000large} leads the application of network models in biological metabolism. Both are among the most novel research according to our metrics: top 8\% for \cite{hidalgo2007product} and top 22\% for \cite{jeong2000large} among the 16 million papers analyzed. We also analyzed a list of Nobel Prize winning papers \cite{li2019dataset}, and confirmed that our metric consistently distinguishes novel from conventional research. We note that Nobel Prize winning papers have a 150\% higher likelihood of being among the top 1\% most novel papers than the average comparable papers in the dataset.

\section{Quantifying the Developmental Index of Research Papers}

The developmental index (Fig.2a) is defined as the inverse of disruption \cite{Wu2019-al}. The Developmental index $D$ measures to what extend a paper incrementally refines previous work rather than radically challenging it, which is computed as,

\begin{align*}
D = \frac{n_b-n_f}{n_f+n_b+n_r}
\end{align*}
where $n_f$ is the number of subsequent works solely cite the focal work, $n_b$ is the number of those cite both the focal work and its references, and $n_r$ is the number of those solely cite its references. Higher $D$ means the paper is more likely to develop existing ideas.

\section{Quantifying the Productivity of Research Papers}
Productivity (Fig.2b) is measured as the total number of papers published by an author in the same year when they contributed to the specific paper under analysis. More publications indicate the author is more productive. We separately analyzed the productivity of lead authors and support authors for each paper (Fig. 2b). 

\section{Quantifying the Impact of Research Papers}
We measure short-term impact $I_{short}$ as the number of citations a paper receives within ten years and long-term impact $I_{long}$ as citations received after twenty years (Fig. 2b), which can be computed as follows:

\begin{align*}
I_{short} = \sum_{i=0}^{10}c_i, \ \ \ \ I_{long} = \sum_{i=21}^{\infty}c_i;
\end{align*}
where $c_i$ is the number of citation each paper receives at the $i$-th year since its publication. Higher $I_{short}$ and $I_{long}$ mean the paper has higher short-term and long-term impact, respectively.

\section{Quantifying the Effect of Public Funding}

External influences such as research funding could have an impact on underlying team structure. For example, research projects supported by large grants may be more likely to have large and hierarchical teams. To explore the influence of research funding, we collect an additional dataset from Clarivate’s Web of Science covering 3,717,823 papers, which labels grant numbers, coupled with funding amounts collected from U.S. National Science Foundation web pages. We find that the percentage of hierarchical teams increased from 48\% to 75\% in papers associated with public funding during 1950~2015, which is more dramatic than the 20\% increases observed in the general pool of scientific articles. This suggests that current funding policies may have an important relationship with (and potentially influence on) team structure. Nevertheless, we find that controlling for variables regarding whether research is funded, and the number and award amount of received grants, L-ratio continues to exercise a significant and substantial independent effect on all six performance metrics. This suggests that even within teams formulated by history or program to be more hierarchical, those that are flatter produce more novel papers, on average.

\section{Verifying L-ratio by Imputed Author Contributions}

We assessed contribution to the focal paper in four distinct ways, which is described as follows. 1) Contribution to references is computed as the overlap between references of the focal paper and all references across previously published papers for each author. It measures to what extent the current references come from a specific author’s “knowledge base”. 2) Contribution to topics is defined as overlap between Microsoft Academic Graph (MAG) topic keywords for the focal paper and all keywords across previously published papers for each author. It measures to what extent the current research topic draws upon a specific author’s previous published work. 3) Contribution in leading the research is calculated as the probability of being the first author(s). 4) Contribution in managing correspondence and presentation is quantified as the probability of being the corresponding author(s). To ensure all calculated metrics were comparable between teams (papers), we applied min-max scaling to convert the individual contribution index calculated in 1) and 2) to between zero and one. 

We calculated contribution to science in four different ways, which is described as follows. 1) Career Age is the number of years from the first publication to the publication of the focal paper for a given author. It approximates the total number of years a scientist has worked as a researcher. 2) Citation is the total number of citations an author has received to all previous publications. It is a proxy to for utility of one’s work to others within the scientific community. 3) Topic Diversity is the total number of unique topic keywords across previous publications and measures the explored knowledge space of a scientist. 4) Publication is the total number of previous papers and measures a scientist’s scientific productivity. We applied min-max scaling to convert all four variables to between zero and one. 

After we obtain a specific contribution index for each author in the focal paper, we calculate three values across all studied papers, including the population average, the average for lead authors, and the average for support authors, before we calculate and display the relative distance from group to population average (Fig. 1b). Bootstrapped 95\% confidence intervals are also calculated and shown in the same figure.

\section{Linkage to Collective Credit Allocation \cite{shen2014collective}}

Shen \& Barabási proposed a citation network-based metric to measure the shifting social perception of credit that the scientific community attributed to each individual author \cite{shen2014collective}. They demonstrate that persistent and focused productivity largely conditions socially perceived credit for past contributions. Our study shows that over historical time, junior scholars are increasingly likely to be involved in taller teams (the fraction of tall teams with an L-ratio below 0.5 increased from 50\% to 70\% over the past five decades), which has enhanced the productivity for those on top and decreased it for those on bottom. This suggests that the shift in credit distribution will likely favor senior researchers who have accumulated more publications over their career. Putting these findings together, the unequal benefits in productivity from increasingly dominant tall teams could exacerbate the unequal distribution of credit for junior scholars. 

To evaluate the correlation between L-ratio and Shen and Barabasi's measure, we map the 30 multi-author Nobel prize winning papers evaluated in Shen and Barabasi’s work to Microsoft Academic Graph (MAG) data, which traces the contribution of 171 authors in total, including 39 Nobel laureates. We find that lead author probability predicted by our model has a moderate and significant Pearson correlation with the collective credit metric (r=0.579, p-value<0.01). In predicting Nobel laureates our proposed lead author probability differs from collective credit for 8 candidates (20.5\% of the 39 Nobel laureates), but our prediction accuracy is very similar, 31 versus 33 successfully predicted Nobel laureates (79.5\% vs. 84.6\% accuracy) from our and their methods, respectively. This demonstrates that L-ratio captures different, complementary signals from collective credit, and that it achieves comparable prediction performance many years before the collective credit assessment has emerged and can be calculated, as it does not rely on post-publication information, i.e., the scientific activity and acclaim of authors in the future. Moreover, we find that Nobel laureates are disproportionately more likely to begin their careers by working in flat teams (L-ratio>0.5) for their first publication with a probability 42.4\% higher than the general population of scientists. These patterns combine to suggest that the L-ratio captures novel signals for long-term scientific impact present within scientist's early stages. They also propose a new potential mechanism through which early mentorship experience influences late career success \cite{malmgren2010role,ma2020mentorship}: flat teams on which young scientists not only ``act for'' their mentors, but ``think with'' them.

\bibliographystyle{unsrtnat}
\bibliography{main}

\end{document}